\documentclass[a4paper,11pt]{article}
\pdfoutput=1 

\usepackage{jcappub}

\usepackage[T1]{fontenc} 
\usepackage{comment}

\usepackage{xcolor,cancel}
\usepackage{orcidlink}
\usepackage{adjustbox}

\title{\boldmath New realisation of light thermal dark matter with enhanced detection prospects}

\author[a]{Amit Adhikary \orcidlink{0000-0001-5269-0021}, }
\author[b,c]{Debasish Borah \orcidlink{0000-0001-8375-282X},}
\author[d]{Satyabrata Mahapatra \orcidlink{0000-0002-4000-5071},}
\author[b]{Indrajit Saha \orcidlink{0000-0002-7459-0838},}
\author[e]{Narendra Sahu \orcidlink{0000-0002-9675-0484},}
\author[e]{and Vicky Singh Thounaojam \orcidlink{0009-0001-6257-5171}}
\affiliation[a]{Aix Marseille Univ., Université de Toulon, CNRS, CPT, IPhU, Marseille, France}
\affiliation[b]{Department of Physics, Indian Institute of Technology Guwahati, Assam 781039, India}
\affiliation[c]{International Institute of Physics, Universidade Federal do Rio Grande do Norte,
Campus Universitario, Lagoa Nova, Natal-RN 59078-970, Brazil}
\affiliation[d]{Department of Physics and Institute of Basic Science, Sungkyunkwan University, 2066 Seobu-ro, Suwon-si, Gyeonggi-do, 16419, Korea}
\affiliation[e]{Department of Physics, Indian Institute of Technology Hyderabad, Kandi, Sangareddy 502285, Telangana, India}

\emailAdd{amit.adhikary@cpt.univ-mrs.fr}
\emailAdd{dborah@iitg.ac.in}
\emailAdd{satyabrata@skku.edu}
\emailAdd{s.indrajit@iitg.ac.in}
\emailAdd{nsahu@phy.iith.ac.in}
\emailAdd{ph22resch01004@iith.ac.in}

\abstract{Light dark matter (DM) with mass around the GeV scale faces weaker bounds from direct detection experiments. If DM couples strongly to a light mediator, it is possible to have observable direct detection rate. However, this also leads to a thermally under-abundant DM relic due to efficient annihilation into light mediators. We propose a novel scenario where a first-order phase transition (FOPT) occurring at MeV scale can restore GeV scale DM relic by changing the mediator mass sharply at the nucleation temperature. The MeV scale FOPT predicts stochastic gravitational waves with nano-Hz frequencies within reach of pulsar timing array (PTA) based experiments like NANOGrav. In addition to enhancing direct detection rate, the light mediator can also give rise to the required DM self-interactions necessary to solve the small scale structure issues of cold dark matter. The existence of light scalar mediator and its mixing with the Higgs keep the scenario verifiable at different particle physics experiments. }

\hypersetup{
colorlinks = true,
linkcolor = blue,
citecolor = magenta
}

\begin{document}
\maketitle
\flushbottom

\section{Introduction}\label{intro}
The matter content of our present universe is dominated by dark matter (DM) contributing approximately five times of ordinary baryonic matter. Weakly interacting massive particle (WIMP) has been the most widely studied candidate for DM which gets produced thermally in the early universe leaving a freeze-out relic. For DM interactions typically in the WIMP ballpark, the requirement of DM not overclosing the universe leads to a lower bound on its mass, around a few GeV \cite{Lee:1977ua, Kolb:1985nn}(with some exceptions for scalar DM \cite{Boehm:2003hm}). On the other hand, light thermal DM with mass $(m_{\rm DM} \lesssim \mathcal{O}(10 \, \rm GeV))$ has received lots of attention in recent times, particularly due to weaker constraints from direct detection experiments like LZ \cite{LUX-ZEPLIN:2022qhg}. 
Such GeV scale DM can have observable direct detection rate if they interact with the SM via a light mediator having sizeable coupling with DM. However, thermal relic of such DM gets under-produced via freeze-out mechanism due to efficient annihilation into light mediators. In this letter, we propose a novel mechanism to revive DM in such thermally under-produced regime via a first-order phase transition (FOPT) occurring at a temperature $T < m_{\rm DM}$, the mass of DM. 
The FOPT leads to sharp change in the mediator mass keeping final relic of DM close to the observed relic \cite{Planck:2018vyg}. 
Such a scenario can not only lead to an observable direct detection rate even in low DM mass regime but can also give rise to self-interacting DM (SIDM), a popular alternative to the cold dark matter (CDM) paradigm solving the small-scale issues like too-big-to-fail, missing satellite and core-cusp problems \cite{Spergel:1999mh, Tulin:2017ara, Bullock:2017xww}.

 We implement our novel production mechanism of thermally under-abundant DM to GeV scale DM with light mediator. 
A FOPT in the early universe assists in generating the correct relic \cite{Planck:2018vyg} of GeV scale thermal DM as follows. DM freezes out by virtue of its coupling to the mediator having similar initial mass as DM, leaving an over-abundant thermal relic prior to the FOPT. The singlet scalar driving the FOPT, by virtue of its coupling to the mediator $\phi$ leads to a sharp fall in the mediator mass after the phase transition. This again leads to DM annihilation into light mediators in such a way that the final relic matches with the observed DM relic. 
In the absence of the FOPT, the thermal relic for same DM and light mediator masses, remains suppressed by more than an order of magnitude compared to the observed limit. 
While the sharp fall in mediator mass still keeps DM relic within observed limits, it enhances the direct detection rate of DM, keeping it verifiable at experiments like CRESST~\cite{Billard:2021uyg} and XENONnT \cite{XENON:2023cxc}. The required mediator mass for GeV scale DM also forces the FOPT to MeV regime where the stochastic gravitational waves (GW) generated by the FOPT have nano-Hz scale peak frequencies which can be probed at pulsar timing array (PTA) based experiments\footnote{In \cite{Han:2023olf}, authors considered self-interacting DM with vector mediator where the latter acquires MeV scale mass from a low scale FOPT. Here, for the first time, we discuss the non-trivial role of FOPT in generating thermal relic of such DM with light mediators.}. In fact, a low scale FOPT provides a good fit \cite{NANOGrav:2023hvm, Megias:2023kiy, Fujikura:2023lkn, Han:2023olf, Zu:2023olm, Athron:2023mer} to the recent data from five different PTA experiments including NANOGrav \cite{NANOGrav:2023gor}.
 In addition to such direct detection and GW signatures, the model has interesting consequences for astrophysical structure formation and remains verifiable at several other experiments like beam dump, collider, cosmic microwave background (CMB) missions etc.

This paper is organised as follows. In section \ref{sec1}, we discuss the scenario of GeV scale DM with strong coupling to a light mediator and our proposed mechanism to revive its thermal relic. In section \ref{sec2}, we discuss the details of detection prospects ranging from gravitational wave, dark matter direct detection, astrophysical structures to other laboratory searches. Finally, we conclude in section \ref{sec3}.

\section{New realisation of light thermal dark matter}
\label{sec1}
\subsection{DM with a light mediator}
We consider a Dirac fermion DM $\chi$ of mass $m_\chi$, singlet under the standard model (SM) gauge symmetry which interacts via a light real scalar mediator $\phi$ as $\mathcal{L}_{\rm DM} \supset \lambda_\chi \overline{\chi} \chi \phi$. The dominant number changing process, deciding the relic density of DM is its annihilation into the light mediator.  
The cross-section for the \( \chi \chi \to \phi \phi \) process is given by:
\begin{equation}
\langle \sigma v\rangle_{\chi\chi\to \phi\phi} = \frac{3}{4} \frac{\lambda^4_\chi}{16 \pi m^2_\chi} v^2 \left(1-\frac{m^2_\phi}{m^2_{\chi}}\right)^{1/2}
\label{Eq:dmann}
\end{equation}
with $m_\chi, m_\phi$ being the masses of DM and mediator respectively. For sufficiently strong DM-mediator interaction, governed by the coupling $\lambda_\chi$, we have under-abundant thermal relic of DM. See appendix \ref{app::SIDM cross-sections} for details. We propose to revive this scenario by a FOPT which changes the mediator mass sharply at nucleation temperature, as we discuss below. \\

\subsection{First-order phase transition}
In addition to fermion dark matter $\chi$ and its mediator $\phi$, we consider another singlet scalar $\phi'$ which drives a FOPT while acquiring a vacuum expectation value (VEV) denoted by $v_{\phi'}$. The Lagrangian relevant for DM and mediator can be written as
\begin{eqnarray}
\label{Lagrangian}
-\mathcal{L}&=& \frac{1}{2} \mu_\phi^{2} \phi^{2}+\mu_{\phi^\prime \phi} \phi^\prime {\phi}^{2} +\frac{1}{4} \lambda_{\phi^\prime \phi} {\phi^\prime}^{2} \phi^{2} + \lambda_\chi \phi \bar{\chi}\chi  \nonumber\\ &+&\frac{1}{2} \lambda_{\phi_1\phi^\prime } {\phi^\prime}^{2} |\phi_1|^{2} + \frac{1}{4}\lambda' \phi'^4\,,
\label{eq1}
\end{eqnarray}
where $\phi_1$ is a scalar-doublet which assists in the FOPT. The mass of the scalar field $\phi$ after the phase transition depends on $v_\phi'$ and can be written as 
\begin{equation}
\label{eq:mphi}
    m_\phi^2=\mu_\phi^2 +2\mu_{\phi^\prime \phi} v_{\phi'} +\frac{\lambda_{\phi^\prime \phi}}{2}v_{\phi'}^2
\end{equation}

Therefore, with suitable choices of parameters, it is possible to decrease the mass of the mediator $\phi$ sharply after the phase transition, compared to its initial mass $\mu_\phi$. Hereafter, we denote the mediator mass after the FOPT as $M_S$. In Eq. \eqref{eq1}, an additional complex scalar field $\phi_1$ is included whose coupling with the scalar field $\phi'$ assists in getting the desired FOPT.

As shown in Appendix \ref{appen2}, we calculate the complete potential including the tree level potential $V_{\rm tree}$, one-loop Coleman-Weinberg potential $V_{\rm CW}$\cite{Coleman:1973jx} along with the finite-temperature potential $V_{\rm th}$ \cite{Dolan:1973qd,Quiros:1999jp}. We also incorporate appropriate decoupling of heavier degrees of freedom \cite{Biondini:2020oib} while studying the low scale FOPT driven by the singlet scalar field $\phi'$. See Appendix \ref{sec:dec_method} for details of the decoupling method. The critical temperature $T_c$ where two degenerate minima $(0, v_c)$ of the potential arise, is estimated by checking the temperature evolution of the potential. The ratio $v_c/T_c$ is identified as the order parameter such that a larger $v_c/T_c$ indicates a stronger FOPT. The FOPT proceeds via tunneling, the rate of which is estimated by calculating the bounce action $S_3$ using the prescription in \cite{Linde:1980tt, Adams:1993zs}. The nucleation temperature $T_n$ is then calculated by comparing the tunneling rate with the Hubble expansion rate of the universe $\Gamma (T_n) = \mathcal{H}^4(T_n) = \mathcal{H}^4_*$. 

We then calculate the relevant parameters required to estimate the stochastic GW spectrum originating from the bubble collisions~\cite{Turner:1990rc,Kosowsky:1991ua,Kosowsky:1992rz,Kosowsky:1992vn,Turner:1992tz}, the sound wave of the plasma~\cite{Hindmarsh:2013xza,Giblin:2014qia,Hindmarsh:2015qta,Hindmarsh:2017gnf} and the turbulence of the plasma~\cite{Kamionkowski:1993fg,Kosowsky:2001xp,Caprini:2006jb,Gogoberidze:2007an,Caprini:2009yp,Niksa:2018ofa}. The two key parameters for GW estimate namely, the duration of the phase transition and the latent heat released relative to radiation density $(\rho_{\rm rad})$ are calculated and parametrised in terms of $\frac{\beta}{{\mathcal{ H}}(T)}$ and $\alpha_*$
\cite{Caprini:2015zlo}
$$\frac{\beta}{{\mathcal{ H}}(T)} \simeq T\frac{d}{dT} \left(\frac{S_3}{T} \right) $$ and 
$$ \alpha_* =\frac{1}{\rho_{\rm rad}}\left[\Delta V_{\rm tot} - \frac{T}{4} \frac{\partial \Delta V_{\rm tot}}{\partial T}\right]_{T=T_n} $$ 
respectively. 
In the expression for $\alpha_*$, the energy difference between true and false vacua is denoted by $\Delta V_{\rm tot}$. 
The bubble wall velocity $v_w$ is estimated from the Jouguet velocity $v_J$\cite{Kamionkowski:1993fg, Steinhardt:1981ct, Espinosa:2010hh}
$$v_J = \frac{1/\sqrt{3} + \sqrt{\alpha^2_* + 2\alpha_*/3}}{1+\alpha_*}$$
according to the prescription outlined in \cite{Lewicki:2021pgr}. The release of latent heat during the FOPT leads to a reheating temperature $T_{\rm RH}$ defined as $T_{\rm RH} = {\rm Max}[T_n, T_{\rm inf}]$ \cite{Baldes:2021vyz} where $T_{\rm inf}$ is determined by $ \rho_{\rm rad}(T_{\rm inf})=\Delta V_{\rm tot}$. Table \ref{tab1} shows a few benchmark points of the model and relevant parameters computed for GW spectrum and desired DM phenomenology. We choose our benchmark points such that scale and strength of phase transition remain safe from CMB and big bang nucleosynthesis (BBN) bounds \cite{Bai:2021ibt}. \\


\begin{table*}
\begin{adjustbox}{width=1.1\columnwidth,center}
    \centering
    \begin{tabular}{|c|c|c|c|c|c|c|c|c|c|c|c|c|c|c|c|}
    \hline
      &  $v_{\phi'}$  & $T_c$  & $v_c$  & $T_n$  & $\mu_\phi$ & $M_S$  & $\mu_{\phi'\phi}$  & $\lambda_{\phi'\phi}$ & $\lambda_{\phi'\phi_1}$ & $\lambda_{\phi_1}$  & $\lambda'$ & $\alpha_*$ & $\beta/\mathcal{H}_* $ & $v_w$ & $T_{\rm RH}$\\
        
        &  (MeV) & (MeV)  &  (MeV) &  (MeV) & (MeV) & (MeV)  & (GeV)   &  & &  &  &  &  &   & (MeV)\\
        \hline
     BP1 & 25 & 7.91 & 23.75 & 3.10 & 438 & 10 & -3.839 & 0.4 & 1.76 & 0.1 &  0.01 & 1.61 & 82.15 & 0.94 & 3.14\\ 
     \hline
     BP2 & 30 & 9.69 & 28.20 & 5.38 & 750 & 15 & -9.377 & 0.3 & 1.70 &  0.1 & 0.01 & 0.39 & 363.32 & 1 & 5.38\\
     \hline
     BP3 & 40 & 12.86 & 38.00 & 5.10 & 990 & 20 & -12.253 & 0.2 & 1.76 &  0.2 & 0.01 & 1.46 & 86.34 & 0.94 & 5.46\\ 
     \hline
     BP4 & 22 & 7.03 & 20.68 & 3.51 & 1200 & 25 & -32.727 & 0.1 & 1.73 &  0.1 & 0.01 & 0.61 & 255.67 & 1 & 3.51\\ 
     \hline
    \end{tabular}
     \end{adjustbox}
    \caption{Benchmark parameters and other details involved in the GW spectrum calculation of the model.}
     \label{tab1}
\end{table*}


\begin{figure}
\centering
\includegraphics[scale=0.5]{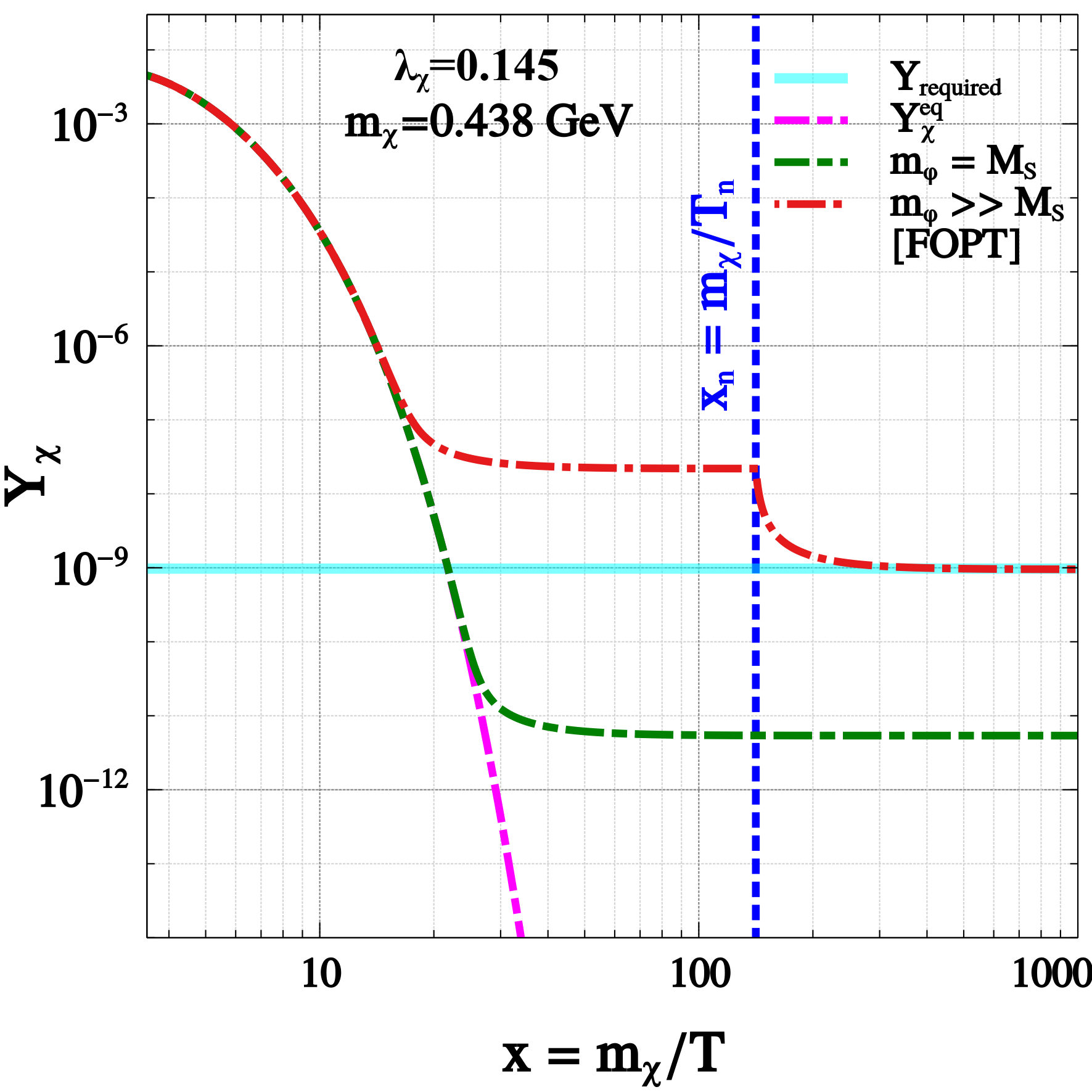}
\includegraphics[scale=0.5]{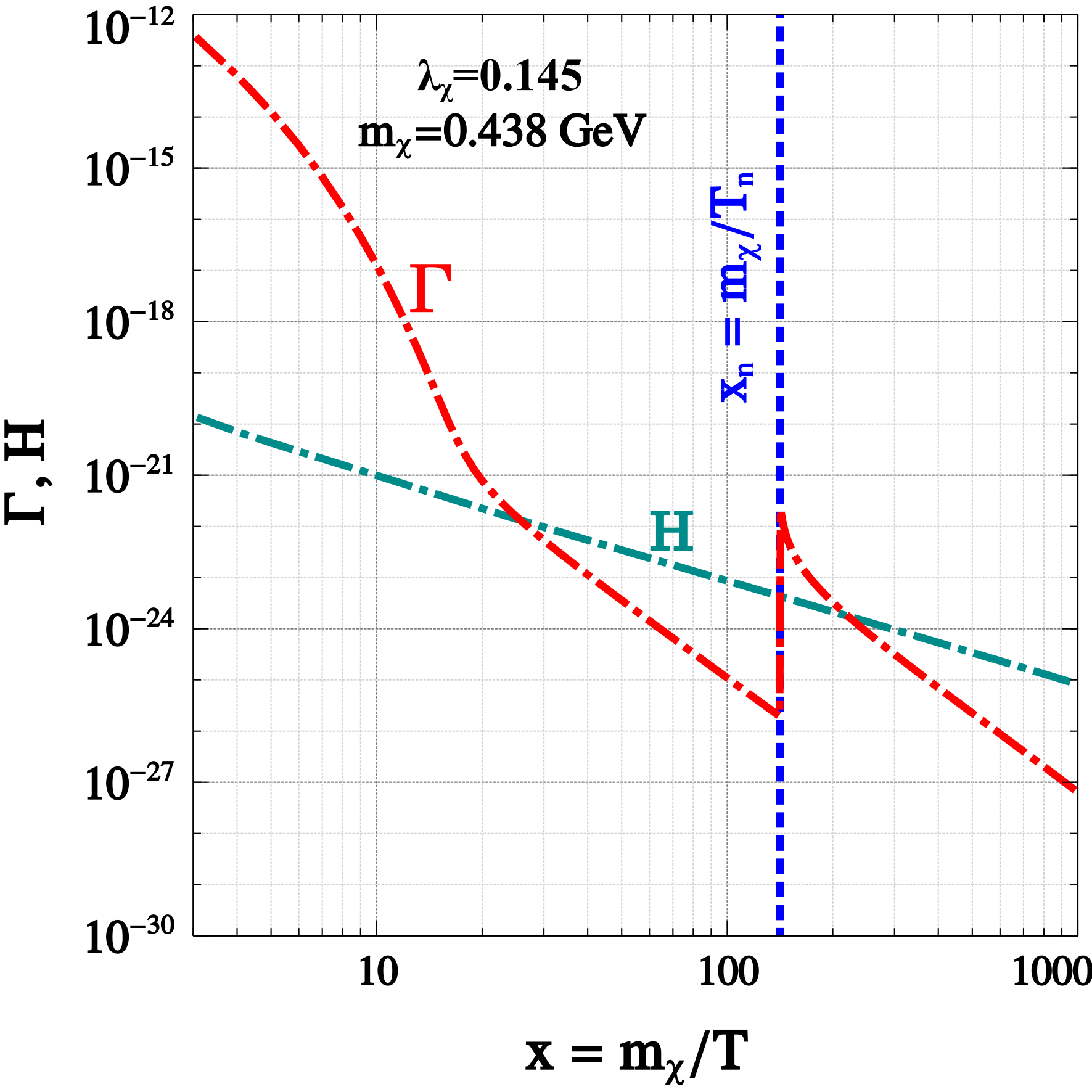}
\caption{Left panel: Evolution of comoving number density of $\chi$ for benchmark values of the parameter. Right panel: Comparison of DM interaction rate with Hubble rate of expansion of the universe.}
\label{fig:plot_allowed}
\end{figure}

\subsection{Light thermal DM with FOPT}
We consider the mediator mass to be same order as DM initially such that the frozen out abundance of DM prior to the nucleation temperature of the FOPT is sufficiently large. In spite of sizeable DM-mediator coupling, one can choose the initial mediator mass to be close to DM mass, either in kinematically allowed or kinematically forbidden regime \cite{DAgnolo:2015ujb, DAgnolo:2020mpt, Griest:1990kh} such that the frozen-out relic is sufficiently large. Below the nucleation temperature, the mediator mass changes sharply, becoming much lighter compared to the DM. This leads to a second phase of DM annihilation, this time to very light mediators, leaving a final abundance same as the observed DM abundance. The primary requirement for this mechanism to yield the correct final DM relic is to have the first freeze-out of DM before the nucleation temperature $T_n$. For a fixed $T_n$, this mechanism can always yield correct DM relic abundance if DM mass $m_\chi \gtrsim 40~ T_n$. Therefore, given the mass of DM, the scale of FOPT gets restricted. For GeV or sub-GeV scale DM, this naturally leads us to the MeV scale FOPT with interesting consequences for PTA based experiments searching for GW.

To demonstrate the plausibility of the scenario, we show the evolution of comoving number density of DM \(Y_{\chi}=n_{\chi}/s\) in  Fig.~\ref{fig:plot_allowed}. For this we numerically solve the Boltzmann equation for the comoving number density of DM.
which can be written as
\begin{equation}
  \frac{dY_{\chi}}{dx} = -\frac{s(m_{\chi}) \langle \sigma v \rangle_{\chi\chi\to\phi\phi}}{x^2 \mathcal{H}(m_{\chi})}\left(Y_{\chi}^2-(Y^{\rm eq}_{\chi})^2\right) \,,  
\end{equation}
where \(\langle \sigma v \rangle_{\chi\chi\to\phi\phi}\) is the thermal averaged annihilation cross-section of DM into light mediators with $x=m_\chi/T$, $s(m_\chi)$ and $\mathcal{H}(m_\chi)$ are the entropy density and Hubble parameter respectively and $Y^{\rm eq}_\chi$ is the equilibrium comoving number density of $\chi$. 
We incorporate the abrupt change of mass of $\phi$ while solving the Boltzmann equation. It should be noted that the cross-section for DM annihilation into light scalar mediator is velocity suppressed. When DM becomes non-relativistic but still remains in equilibrium, it maintains a Maxwell-Boltzmann distribution of velocities and its mean velocity can be approximated by the thermal velocity $3 T/m_\chi$. After freeze-out, its velocity redshifts as $v\sim a^{-1}$ with $a$ being the scale factor. Hence the velocity of DM at any temperature $T$ post freeze-out can be estimated by $v_\chi=v_f (T/T_f)$, where $T_f$ is the freeze-out temperature of DM and $v_f$ is the corresponding velocity.

From the left panel of Fig.~\ref{fig:plot_allowed}, it is evident that if the mediator initially has a very light mass (i.e., \(m_\phi=10\) MeV), excessive annihilation of DM into the light mediator leads to DM under-abundance, as indicated by the green dot-dashed line. However, by reducing the annihilation rate through adjusting the mediator mass close to the DM mass (\(m_\phi \sim m_\chi \)), thereby reducing the phase space, it is possible to induce early DM freeze-out, resulting in a larger DM freeze-out abundance. This abundance is subsequently depleted after the FOPT due to sudden decrease in mediator mass, which is marked by the blue dashed vertical line in the figure corresponding to BP1 of the Table~\ref{tab1} with \(T_n=3.10\) MeV. 
After FOPT, the depletion of \(Y_\chi\) occurs because the mediator mass abruptly changes to a smaller value during FOPT, breaking the inhibition due to phase space suppression. Eventually, this process falls below the Hubble expansion, saturating the DM abundance to the required value, as depicted by the red dot-dashed line. 
The release of latent heat from the FOPT does not cause much dilution of DM relic as $T_{\rm RH} \approx T_n$ for the benchmark points in Table \ref{tab1}. 
The evolution can also be understood by comparing the interaction rate for the DM annihilation (\(\Gamma=n_\chi \langle \sigma v \rangle_{\chi\chi \to \phi \phi}\)) with the Hubble expansion rate which is shown on the right panel of Fig.~\ref{fig:plot_allowed}.
It is worth mentioning that some earlier works \cite{Elor:2021swj, Baker:2016xzo, Cohen:2008nb, Croon:2020ntf, Hashino:2021dvx} also studied the role of phase transitions on DM properties. But here we utilise it to thermally under-abundant GeV scale DM with enhanced detection prospects ranging from direct detection, astrophysical structure formation, stochastic GW among others.

\section{Detection Prospects}
\label{sec2}

\subsection{Dark matter self-interactions}
A fermion dark matter with Yukawa couplings to a light scalar offers a resolution to the small-scale issues of the \( \Lambda \)CDM model. With such a light mediator, the non-relativistic self-interaction of DM arises due to Yukawa-type potential: \( V(r) = \pm \frac{\alpha_\chi}{r} e^{-m_{\phi}r} \), where \( \alpha_\chi = \frac{\lambda^2_\chi}{4\pi} \) signifies the dark fine structure constant. Depending on the masses of DM (\( m_{\chi} \equiv m_{\rm DM} \)) and the mediator (\( m_{\phi} \)), as well as the relative velocity of DM (\( v \)) and the interaction strength (\( \lambda_{\chi} \)), three distinct regimes emerge: the Born regime (\( \frac{\lambda^2_{\chi} m_{\rm \chi}}{4\pi m_{\phi}} \ll 1, \frac{m_{\rm \chi} v}{m_{\phi}} \geq 1 \)), the classical regime (\( \frac{\lambda^2_{\chi} m_{\rm \chi}}{4\pi m_{\phi}} \geq 1 \)), and the resonant regime (\( \frac{\lambda^2_{\chi} m_{\rm \chi}}{4\pi m_{\phi}} \geq 1, \frac{m_{\rm \chi} v}{m_{\phi}} \leq 1 \))~\cite{Tulin:2013teo,Tulin:2012wi,Khrapak:2003kjw}. The details are given in Appendix \ref{app::SIDM cross-sections}. The desired self-interaction of DM is then parametrised in terms of cross-section to mass ratio as $\sigma/m \sim 1 \; {\rm cm}^2/{\rm g} \approx 2 \times 10^{-24} \; {\rm cm}^2/{\rm GeV}$ \cite{Buckley:2009in, Feng:2009hw, Feng:2009mn, Loeb:2010gj, Zavala:2012us, Vogelsberger:2012ku}.

DM with strong coupling to a light mediator behaves like CDM on large scales while solving the small-scale issues due to its velocity-dependent self-interaction cross-section \cite{Buckley:2009in, Feng:2009hw, Feng:2009mn, Loeb:2010gj, Bringmann:2016din, Kaplinghat:2015aga, Aarssen:2012fx, Tulin:2013teo}. However, the existence of such light mediator having sizeable interactions with DM also leads to under-abundant thermal relic, particularly in the GeV scale mass regime, due to efficient DM annihilation into light mediators. 
While there exist several SIDM production mechanisms in the literature \cite{Kouvaris:2014uoa, Bernal:2015ova, Kainulainen:2015sva, Hambye:2019tjt, Cirelli:2016rnw, Kahlhoefer:2017umn, Belanger:2011ww}, 
Generating correct DM relic purely from thermal freeze-out is challenging, except for a few scenarios where non-standard cosmology \cite{Ho:2023ctb} and conversion driven freeze-out \cite{Borah:2022ask} were adopted. On the other hand, there are several works where a hybrid setup of thermal and non-thermal contributions to relic density was studied \cite{Dutta:2021wbn, Borah:2021pet, Borah:2021rbx, Borah:2021qmi}. In the present proposal, thermal freeze-out is revived for such DM with the help of a first order phase transition, as discussed above.

\begin{figure}[h]
    \centering
    \includegraphics[scale=0.55]{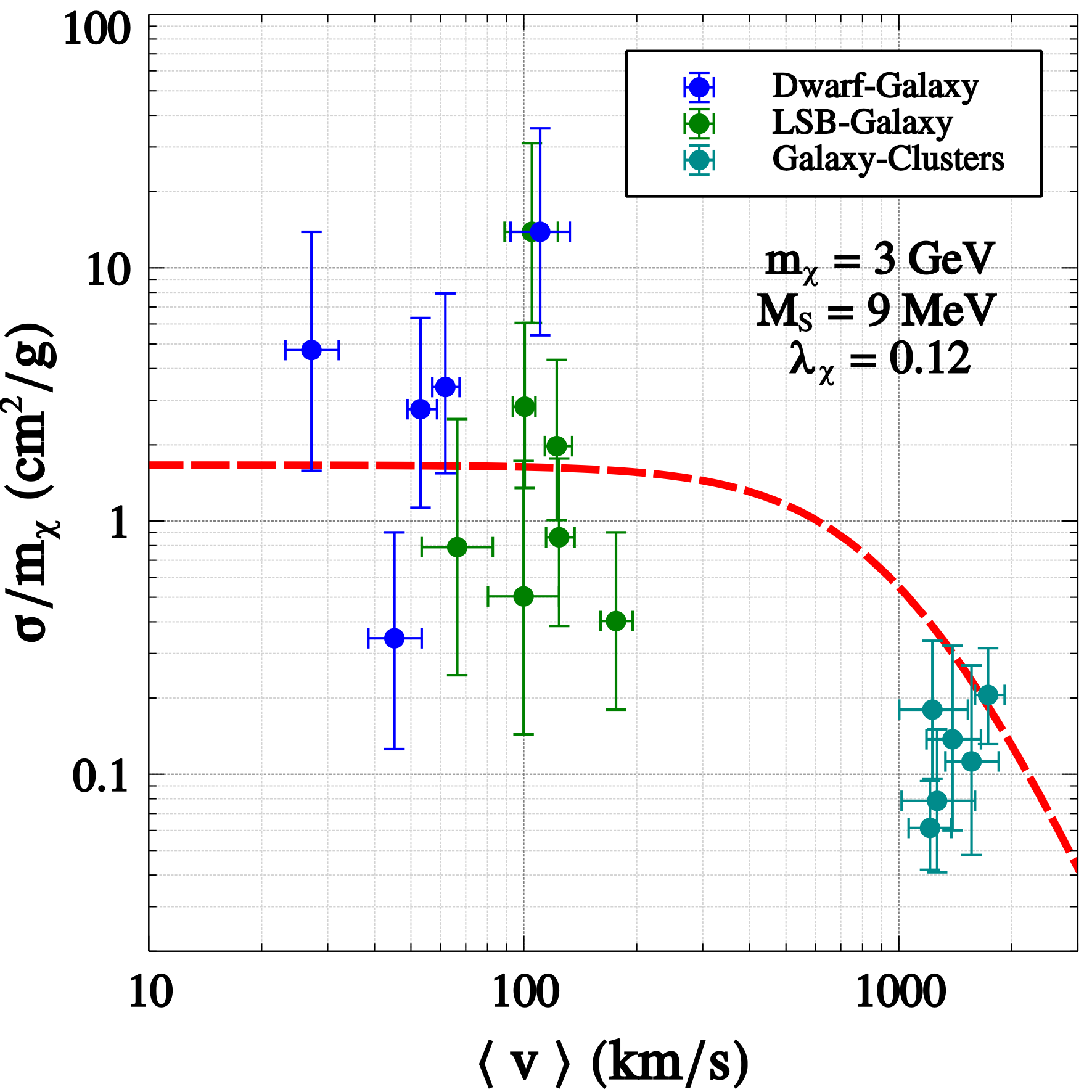}
    \caption{The self-interaction cross-section per unit mass of DM as a function of average collision velocity.}
    \label{fig:sidm_v_profile}
\end{figure}

While our primary motivation was to enhance the direct detection prospects of GeV scale DM while being consistent with observed relic in a minimal setup, the velocity-dependent DM self-interactions come as a bonus. Fig.~\ref{fig:sidm_v_profile} depicts the self-scattering cross-section per unit DM mass as a function of average collision velocity, with data points from dwarfs(blue), low surface brightness (LSB) galaxies (green) and clusters (dark cyan)~\cite{Kamada:2020buc,Kaplinghat:2015aga}. The red dashed curve represents the velocity-dependent cross-section for a benchmark point as mentioned in the inset of the figure that fits well with the astrophysical observations. Due to interesting impact of such DM on astrophysical structure formation, future astrophysical surveys will be able to shed more light into its properties.

\begin{figure}[h]
    \centering
    \includegraphics[scale=0.5]{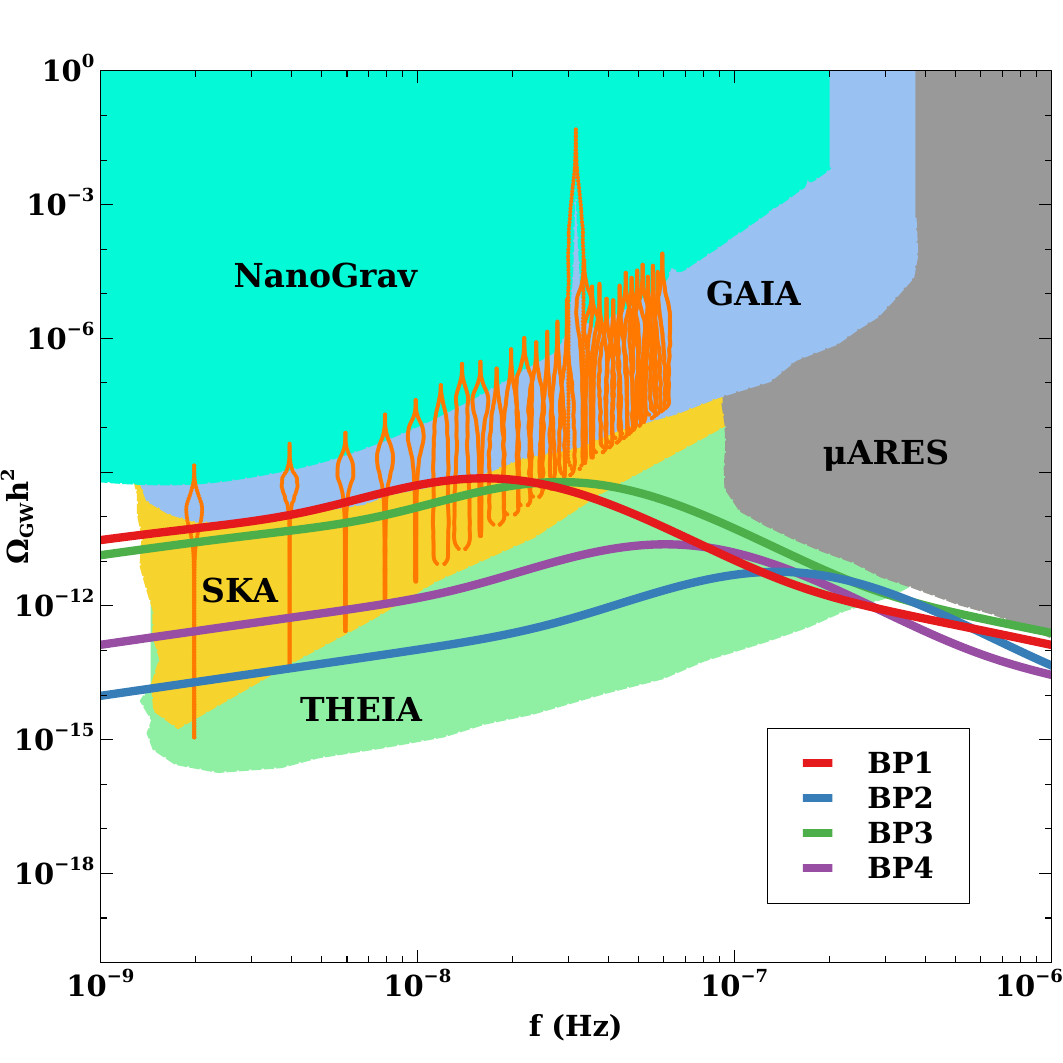}
    \caption{GW Spectrum for the benchmark points given in Table \ref{tab1}.}
    \label{fig:gw}
\end{figure}

\subsection{Gravitational waves}
Low scale phase transition leads to several detection prospects of the model at particle physics, cosmology as well as gravitational wave experiments. We compute the GW spectrum for FOPT in our model following the prescription outlined in Appendix \ref{appen3}. Fig. \ref{fig:gw} shows the GW spectrum for the benchmark points given in Table \ref{tab1}. The sensitivities of NANOGrav \cite{NANOGrav:2023ctt}, GAIA \cite{Garcia-Bellido:2021zgu}, THEIA~\cite{Garcia-Bellido:2021zgu}, $\mu$ARES~\cite{Sesana:2019vho} and SKA~\cite{Weltman:2018zrl} are shown as different coloured shades. The range of GW spectrum from NANOGrav results~\cite{NANOGrav:2023gor} is shown by the orange binned points.  The region above $\Omega_{\rm GW} h^2 \gtrsim 10^{-6}$ is disfavored by the limit on $\Delta N_{\rm eff}$ \cite{Planck:2018vyg} from GW overproduction. Clearly, two of the benchmark points provide a significant overlap of the blue-tilted part of the spectrum with the NANOGrav data. 
While the PTA signal can, in principle, be generated by supermassive black hole binary (SMBHB) mergers though with a mild tension, presence of exotic new physics like a low scale FOPT discussed here provide a good fit \cite{NANOGrav:2023hvm, EPTA:2023xxk}. While the blue-tilted part of the spectrum can give a good fit to the PTA data, the other part of the spectrum can be probed at future experiments sensitive to higher frequencies, keeping the FOPT origin of the PTA signal distinguishable from other new physics origins.

\subsection{Direct Detection}
Owing to the mixing of mediator and SM Higgs, the direct detection of the dark matter through Higgs portal can be realised and the spin-independent (SI) scattering cross-section of DM per nucleon is given by:
 \begin{equation}\label{eq7}
 	\sigma_{\rm SI}=\frac{\mu_r^2}{4\pi A^2}[Zf_p + (A-Z)f_n]^2
 \end{equation}
where $\mu_r$ is the reduced mass of the DM-nucleon system. $A$ and $Z$ are the mass number and the atomic number of the target nucleus respectively. The interaction strengths $f_p$ and $f_n$ \cite{Hoferichter:2017olk} of proton and neutron with DM are given as
 \begin{equation}\label{eq8}
 	f_{p,n}=\sum_{q=u,d,s} f_{T_q}^{p,n}\alpha_q \frac{m_{p,n}}{mq} + \sum_{q=c,t,b} f_{TG}^{p,n}\alpha_q \frac{m_{p,n}}{mq}
 \end{equation}
 with $\alpha_q$ defined by 
 \begin{equation}
 	\alpha_q=\lambda_\chi \sin\theta_{\phi H}\frac{m_q}{v_{\rm EW}}\left[ \frac{1}{M_S^2}-\frac{1}{M_H^2}\right].
 \end{equation}\label{eq9}

\begin{figure}[h]
    \centering
    \includegraphics[scale=0.55]{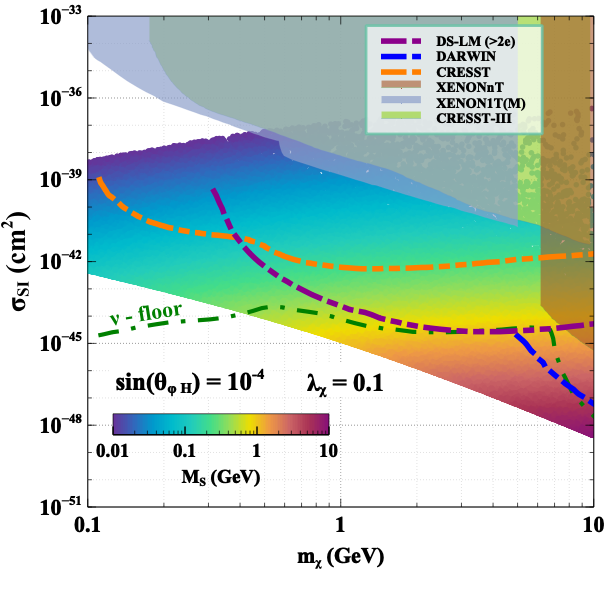}
    \caption{Parameter space showing enhanced direct detection rate for DM-nucleon scattering due to light mediator.}
    \label{fig:dd_constraint}
\end{figure}

The plot in Fig. \ref{fig:dd_constraint} shows the variation of spin-independent WIMP-nucleon cross-section with dark matter mass $m_\chi$. Three of the most stringent constraints from current DM direct detection experiments, applicable in the low DM mass range, are shown as shaded regions. These include CRESST-III~\cite{CRESST:2019jnq}, XENON1T-Migdal~\cite{XENON:2019zpr} and XENONnT~\cite{XENON:2023cxc}. We take benchmark values of $\lambda_\chi = 0.1$ and ${\rm sin}(\theta_{\phi H}) = 10^{-4}$ and calculate the spin-independent scattering cross-section by varying the mediator mass from $10$ MeV to values close to $m_\chi$. The result is shown by color-shaded region. It becomes evident that for a given $m_\chi$, a smaller mediator mass results in an enhanced scattering cross-section, which can either be ruled out by current bounds or remain within future sensitivity projections (90\% CL). These projections are indicated by the dashed-dotted lines and include those from CRESST~\cite{Billard:2021uyg}, DARWIN~\cite{DARWIN:2016hyl, Schumann:2015cpa} and DS-LM~\cite{GlobalArgonDarkMatter:2022xgs}.

\begin{figure}[h]
    \centering
    \includegraphics[scale=0.55]{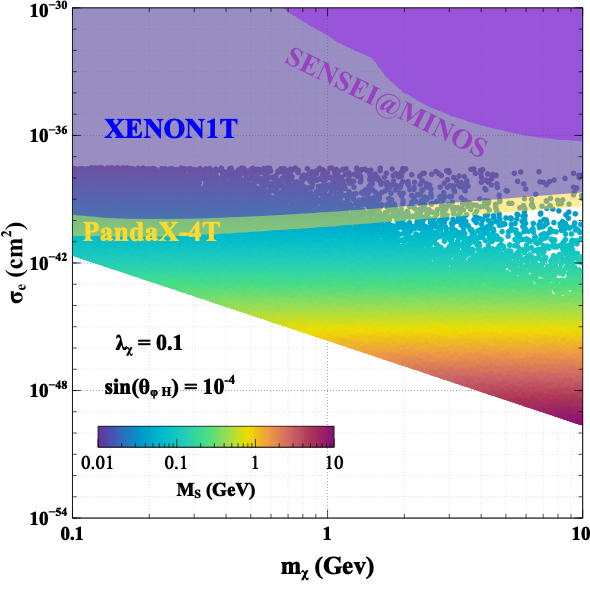}
    \caption{Parameter space showing enhanced direct detection rate for DM-electron scattering due to light mediator.}
    \label{fig:dd_electron}
\end{figure}

Constraints from DM-electron scattering is also shown in Fig.~\ref{fig:dd_electron} against the current bound from XENON1T~\cite{XENON:2019gfn}, SENSEI$@$MINOS~\cite{SENSEI:2020dpa} and PandaX-4T~\cite{PandaX:2022xqx} in the plane of $\sigma_e$ versus $m_\chi$. The color-shaded region is again plotted by following the aforementioned benchmark. The formula to calculate the DM-electron scattering cross-section is given by
\begin{equation}
    \sigma_e = \frac{16 \pi \mu_{\chi e}^2 \alpha \alpha_\chi {\rm sin(\theta_{\phi H})^2}}{M_S^4}~{|F(q)|^2}
\end{equation}
where $\alpha_{\chi}=\lambda_\chi^2/4 \pi$ and $\mu_{\chi e}$ is the DM-electron reduced mass and $F(q)=1$ for a massive mediator and $\alpha^2 m_e^2/q^2$ for a massless mediator.

\begin{figure}[h]
    \centering
    \includegraphics[scale=0.55]{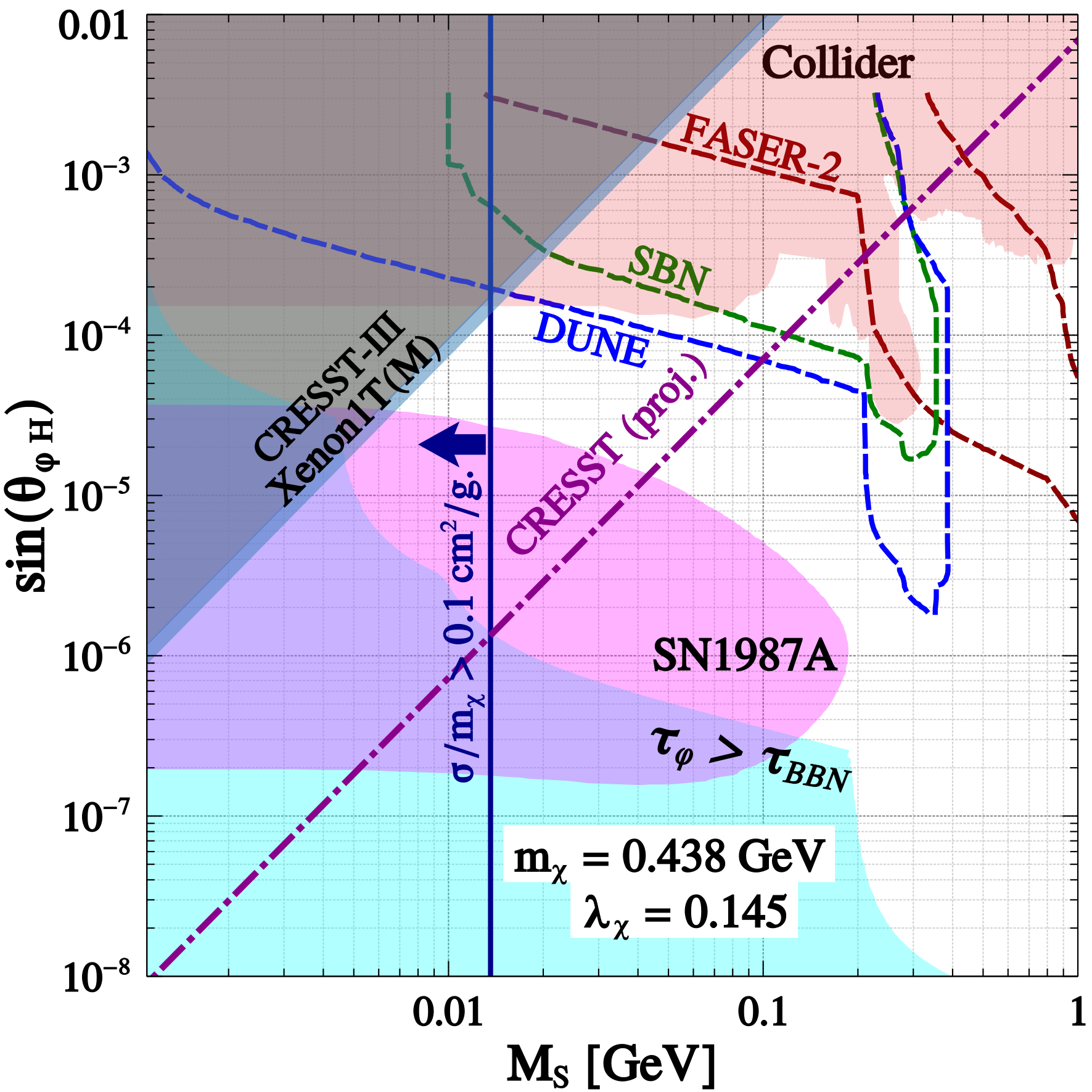}
    \caption{Parameter space in the plane of singlet-Higgs mixing $\theta_{\phi H}$ and mediator mass $M_S$ showing relevant constraints and experimental sensitivities.}
    \label{fig:summary}
\end{figure}

We summarise our results in Fig. \ref{fig:summary} showing the parameter space in the plane of mediator mass versus mediator-Higgs mixing $\theta_{\phi H}$. Such a mixing can arise due to a trilinear term of the form $\mu_{\phi H} \phi H^\dagger H$. 
While a tiny mixing of $\phi$ with the SM Higgs plays no role in the FOPT as well as DM relic, such a mixing is crucial to ensure that $\phi$ is not cosmologically long-lived and decays into the SM particles before the BBN epoch. A non-zero mixing with the Higgs also brings complementary detection prospects as indicated in Fig. \ref{fig:summary}. 
The red shaded region represents constraints from collider searches, specifically those arising from flavor-changing neutral current (FCNC) meson decays~\cite{Dev:2019hho,Dev:2017dui} combining the constraints from  NA48/2~\cite{NA482:2010zrc}, E949~\cite{BNL-E949:2009dza}, KOTO~\cite{KOTO:2018dsc}, NA62~\cite{NA62:2014ybm}, KTeV~\cite{KTeV:2008nqz}, BaBar~\cite{BaBar:2014zli}, Belle~\cite{Belle:2009zue}, LHCb~\cite{LHCb:2016awg}. Additionally, constraints from the LSND and CHARM experiments are also included, as the particle $\phi$ can be produced through proton bremsstrahlung in the LSND~\cite{Foroughi-Abari:2020gju}, and might also be produced in the high-intensity beamdump experiments like CHARM~\cite{CHARM:1985anb}. The dashed blue, green, and red lines indicate the future detection prospects of DUNE~\cite{Dev:2021qjj}, SBN~\cite{Batell:2019nwo}, and FASER-2~\cite{Anchordoqui:2021ghd}, respectively. The magenta shaded region indicates constraints from SN1987A~\cite{Dev:2020eam}. 
These constraints arise because if such light bosons are abundantly produced and escape from the source, they would conflict with the observed neutrino luminosity. 
Additionally, we have constrained the lifetime of 
$\phi$ to be less than the typical BBN epoch to avoid disturbing the predictions of light nuclei abundance by entropy injection or by changing neutrino decoupling temperature. This constraint disfavours the cyan shaded region. The constraints from the DM direct detection experiments CRESST-III~\cite{CRESST:2019jnq} and XENON1T-Migdal~\cite{XENON:2019zpr} are shown by the grey and dark blue shaded regions for a fixed DM mass $m_\chi=0.438$ GeV and coupling $\lambda_\chi=0.145$. For the same benchmark value of $m_\chi$ and $\lambda_\chi$, we also illustrate the anticipated sensitivity of CRESST~\cite{Billard:2021uyg} by the magenta dot-dashed line which can probe a crucial region of the parameter space. Due to p-wave suppressed annihilation cross-section, our scenario remains safe from the CMB constraints on DM annihilation during the recombination epoch as well as from the indirect DM search experiments~\cite{Hambye:2019tjt}. The region towards the left of the vertical blue line is consistent with the required self-interactions of DM. If DM self-interaction is not required to solve the small-scale structure issues, then the entire white region will be allowed. Additionally, the constraints from direct detection will become weaker opening up more allowed parameter space, as smaller values of DM-mediator coupling get allowed.
\\

\section{Conclusion} 
\label{sec3}
We have proposed a novel scenario for generating correct relic of GeV scale dark matter having enhanced direct detection rate at direct detection experiments like CRESST and XENONnT due to its strong interaction with a light mediator. We start with a mediator mass close to DM mass such that thermal relic of DM remains large initially. Subsequently, a first-order phase transition occurs with a nucleation temperature sufficiently below the initial freeze-out temperature of DM. This leads to a sharp fall of mediator mass thereby enhancing the direct detection rate. This also leads to the second phase of DM annihilation into light mediators leaving a final relic same as the observed DM relic. Keeping DM mass around the GeV scale, naturally forces the FOPT to be in the MeV or few tens of MeV regime with the consequence of generating stochastic GW having peak frequencies in the nano-Hz ballpark. While the blue-tilted part of the GW spectrum can naturally explain the recent PTA data, future experiments can further confirm the FOPT origin of the PTA data. Due to the existence of a light mediator, the model also remains verifiable at different particle physics experiments, thanks to the scalar portal mixing. Additionally, GeV scale fermion dark matter with light scalar mediator can give rise to sufficient DM self-interactions at low energy with the potential to solve the small-scale structure issues of cold dark matter.

\acknowledgments
A.A. received support from the French government under the France 2030 investment plan, as part of the Initiative d'Excellence d'Aix-Marseille Université - A*MIDEX. S.M. acknowledges the financial support from National Research Foundation(NRF) grant funded by the Korea government (MEST) NRF-2022R1A2C1005050.
The work of D.B. is supported by the Science and Engineering Research Board (SERB), Government of India grants MTR/2022/000575, CRG/2022/000603 and by the Simons Foundation
(Award Number:1023171-RC). The works of N.S. is supported by the Department of Atomic Energy- Board of Research in Nuclear Sciences, Government of
India (Ref. Number: 58/14/15/2021- BRNS/37220).

\appendix
\section{Dark matter self-interactions due to light mediator}\label{app::SIDM cross-sections}
The non-relativistic DM self-scattering can be effectively described using the attractive Yukawa potential:
 \begin{equation}
     V(r)=-\frac{\lambda_{\chi}^2}{4\pi r}e^{-M_{S}r}
 \end{equation}
 To capture the relevant physics of forward scattering, the transfer cross-section is defined as
 \begin{equation*}
     \sigma_T=\int d\Omega (1-cos\theta)\frac{d\sigma}{d\Omega}
 \end{equation*}
 In the Born limit, where ${\lambda_{\chi}}^2 m_\chi/(4\pi {M_S})\ll 1$, the transfer cross-section is given by:
 \begin{equation*}
     \sigma^{\rm Born}_T=\frac{{\lambda_{\chi}}^4}{2\pi {m_\chi^2} v^4} \left[ {\rm log}\left( 1+\frac{m^2_{\chi} v^2}{M_S^2} \right)-\frac{m_\chi^2 v^2}{M_S^2+m_\chi^2 v^2} \right]
 \end{equation*}
 
 Outside the Born limit, where $ \lambda^2_{\chi} m_\chi/(4\pi M_S)\geq 1$, there are two different regimes: the classical regime and the resonance regime. In the classical regime ($m_\chi/M_S \geq 1$), solution for an attractive potential is given by
     \begin{equation*}
   \sigma^{\rm classical}_T=\begin{cases}
     \frac{4\pi}{M_S^2}\beta^2 \ln(1+\beta^{-1}) & \textbf{$\beta > 1$}\\
     \frac{8\pi}{M_S^2}\left[ {\beta^2/(1+1.5\beta^{1.65})} \right] & \textbf{$10^{-1} < \beta \leq 10^3$} \\
     \frac{\pi}{M_S^2}\left[ \ln\beta +1-1/2 \ln^{-1}{\beta}\right]^2 & \textbf{$\beta \geq 10^3$}
   \end{cases}
 \end{equation*}
 where $\beta=\frac{2 \lambda_{\chi}^2{M_{S}}}{4\pi {m_\chi}v^2}$. 

 Finally in the resonance regime ($ m_\chi v/M_S \leq 1$), there is no analytical formula for $\sigma_T$. So approximating the Yukawa potential by Hulthen potential $\left(V(r)=- \frac{{\lambda_{\chi}}^2}{4\pi}\frac{\delta e^{-\delta r}}{1-e^{-\delta r}}\right)$, the transfer cross-section is obtained as: 
     \begin{equation*}
         \sigma_T^{\rm Hulthen}=\frac{16\pi \sin^2\delta_0}{m_\chi^2v^2}
     \end{equation*}
 where $l=0$ phase shift $\delta_0$ is given by:
 $$ \delta_0={\rm Arg} \left[ \frac{i\Gamma(i m_\chi v/kM_S)}{\Gamma(\lambda_+)\Gamma(\lambda_-)} \right]$$
 {\rm with}
 $$\lambda_{\pm}=1+\frac{i m_\chi v}{2kM_S}\pm \sqrt{\frac{{\lambda_{\chi}}^2 m_\chi}{4\pi kM_S}-\frac{m_\chi^2{v^2}}{4k^2 {M^2_S}}}$$
 and $k \approx 1.6$ is a dimensionless number. 

\begin{figure}[h]
    \centering    \includegraphics[scale=0.55]{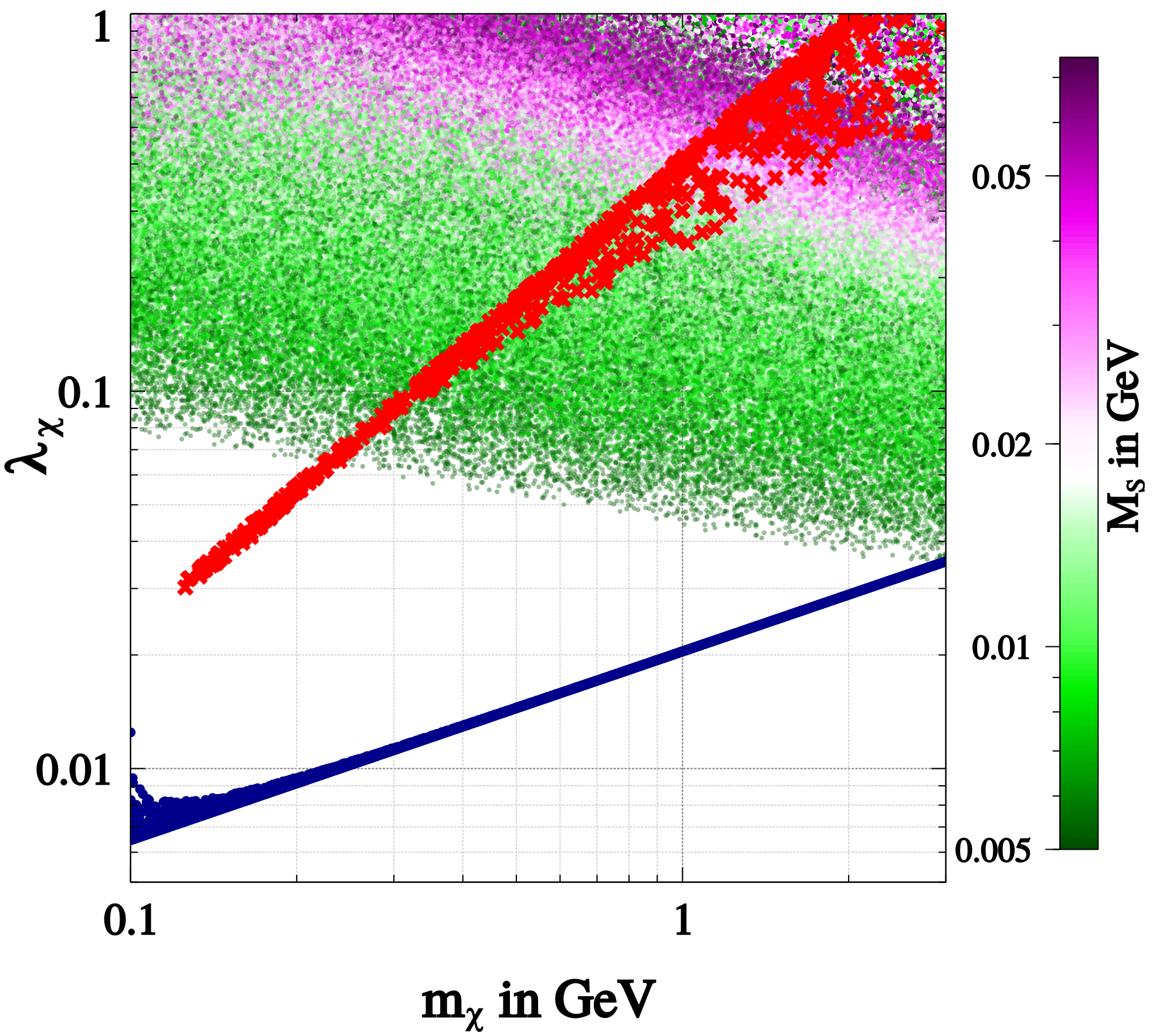}
    \caption{Parameter space in the $\lambda_\chi-m_\chi$ plane consistent with DM self-interaction (colored regions) and correct DM relic without FOPT (blue points) and with FOPT (red) points.}
\label{fig:sidm_relic_ps}
\end{figure}

In Fig.~\ref{fig:sidm_relic_ps}, we illustrate the parameter space that can produce the required self-interaction in the plane of $\lambda_\chi$ and $m_\chi$, while varying the mediator mass $M_S$ in the range of $[5, 75]$ MeV, as indicated by the color code.
The blue points represent the parameter space consistent with the thermal relic DM when there is no FOPT and the mediator mass remains very light throughout cosmological evolution. For GeV and sub-GeV scale DM, this parameter space leads to an under-abundant thermal relic, indicating that no parameter space simultaneously satisfies both the correct relic density and the required DM self-interaction.

In the same plane, the red points indicate the parameter space that meets the correct DM relic density when the mediator mass is close to the DM mass before FOPT and abruptly becomes light after FOPT. This scenario shows a parameter space that is consistent with both the required self-interaction and the correct thermal relic criteria.

\section{First-order phase transition}
\label{appen2}

\begin{figure*}
    \centering
    \includegraphics[scale=0.42]
    {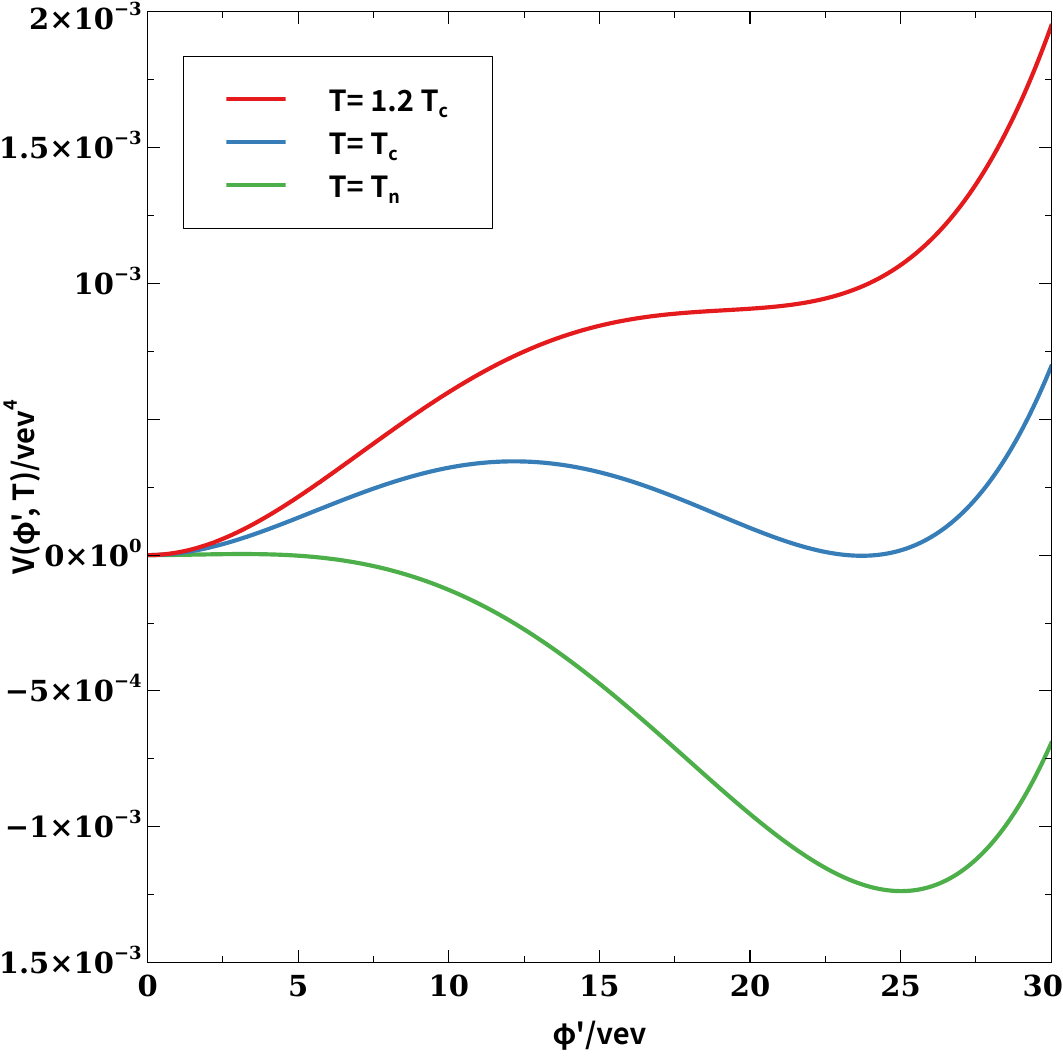}
        \includegraphics[scale=0.5]{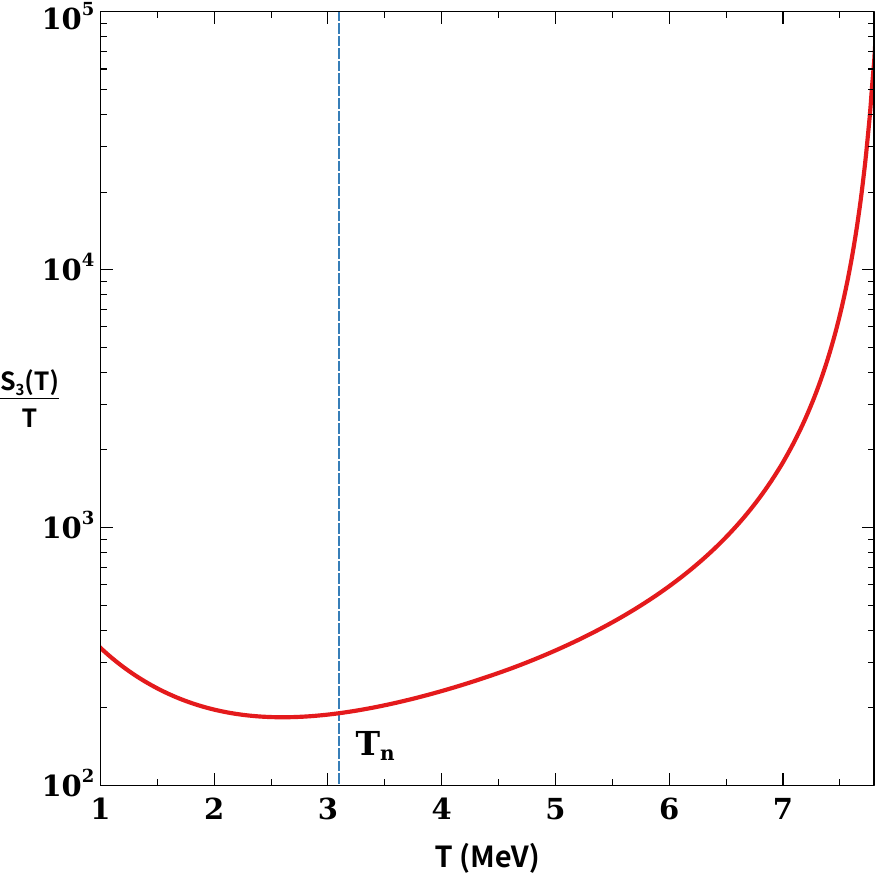}
    \caption{Left panel: Potential profile for benchmark point BP1 given in Table-I of the draft. Right panel: $S_3$/T versus T profile for benchmark point BP1. The vertical dashed line indicates the nucleation temperature.}
    \label{fig:potential}
\end{figure*}

The tree-level scalar potential, relevant for the FOPT, is 
\begin{equation}
    V_{\rm tree}(\phi')=\frac{\lambda'}{4}(\phi'^2-v^2_{\phi'})^2 + \frac{1}{2} \lambda_{\phi_1\phi^\prime } {\phi^\prime}^{2} |\phi_1|^{2}.
\end{equation}
The field dependent masses are
\begin{equation}
    m_{\phi'}^2(\phi')=3\lambda' \phi'^2-\lambda' v^2_{\phi'},  \quad n_{\phi'}=1
\end{equation}
\begin{equation}
    m_{\phi_1}^2(\phi')=\frac{\lambda_{\phi_1 \phi'}}{2}\phi'^2,  \quad n_{\phi_1}=2
\end{equation}

\begin{align}\nonumber
\Pi_{\phi'}(T) &= \left(\frac{\lambda'}{4}+ \frac{\lambda_{\phi' \phi_1}}{6} \right) T^2,\\
\Pi_{\phi_1}(T) &= \left(\frac{\lambda_{\phi_1}}{3}+ \frac{\lambda_{\phi' \phi_1}}{12} \right) T^2,
\end{align} 
where, $\lambda_{\phi_1}$ is the self quartic coupling of $\phi_1$.
The effective potential
\begin{align}
V_{\rm tot} = V_{\rm tree} + V_{\rm CW} +V_{\rm th} +V_{\rm ct}.
\label{eq:potential}
\end{align}
The One-loop Coleman Weinberg potential
\begin{align}
V_{\rm CW} = \sum_i (-)^{n_{f}} \frac{n_i}{64\pi^2} m_i^4 (\phi') \left(\log\left(\frac{m_i^2 (\phi')}{\mu^2} \right)-\frac{3}{2} \right),
\end{align}
The decoupled CW potential for heavy $\phi$ field (Here scale, $\mu=v_{\phi'}\equiv {\rm VEV \, of } \, \phi'$) : 
\begin{align}
V_{\rm CW} = \frac{1}{64\pi^2}~\theta(v_{\phi'} - m_\phi)~m_\phi^4 (\phi') \left(\log\left(\frac{m_\phi^2 (\phi')}{v^2_{\phi'}} \right)-\frac{3}{2} \right).
\label{VCW_deco}
\end{align}
Thermal contribution to effective potential is
\begin{align}
V_{\rm th} = \sum_i \left(\frac{n_{\rm B_i}}{2\pi^2}T^4 J_B \left[\frac{m_{\rm B_i}}{T}\right] - \frac{n_{\rm F_{i}}}{2\pi^2}T^4 J_F \left[\frac{m_{\rm F_{i}}}{T}\right]\right),
\end{align}
where $n_{B_i}$ and $n_{F_i}$ denote the dof of the bosonic and fermionic particles, respectively. The functions $J_B$ and $J_F$ are defined as
\begin{align}
&J_B(x) =\int^\infty_0 dz z^2 \log\left[1-e^{-\sqrt{z^2+x^2}}\right] \label{eq:J_B},\\
&J_F(x) =   \int^\infty_0 dz z^2 \log\left[1+e^{-\sqrt{z^2+x^2}}\right].
\end{align}
The thermal potential including daisy contributions $V_{\rm daisy}$ is
\begin{align}
    V_T(\phi',T) &= V_{\rm th} + V_{\rm daisy}(\phi',T), \\
    V_{\rm daisy}(\phi',T) &= -\sum_i \frac{g_i T}{12\pi}\left[ m^3_i(\phi',T) - m^3_i(\phi') \right]. \nonumber
\end{align}
In Eq. \eqref{eq:potential}, $V_{\rm ct}$ includes the counter-terms. The shape of the effective potential for a benchmark point is shown in Fig. \ref{fig:potential} (left panel) indicating the appearance of two degenerate minima at a critical temperature.

The FOPT then proceeds via tunneling. The tunneling rate can be estimated from the bounce solution. The tunneling rate per unit volume is defined as
\begin{align}
\Gamma (T) = \mathcal{A}(T) e^{-S_3(T)/T}.
\end{align}
where $\mathcal{A}(T)\sim T^4\left( \frac{S_3(T)}{2\pi T}\right)^{3/2}$ and the bounce action $S_3(T)$ are respectively determined by the dimensional analysis and given by the classical configuration, called bounce. The effective potential is fitted to a generic potential for which the action calculations are done in semi-analytical way following the prescription given in \cite{Adams:1993zs}. The right panel of fig. \ref{fig:potential} shows the profile of $S_3/T$ for a chosen benchmark point BP1. 

\begin{figure*}
    \centering
    \includegraphics[scale=0.4]{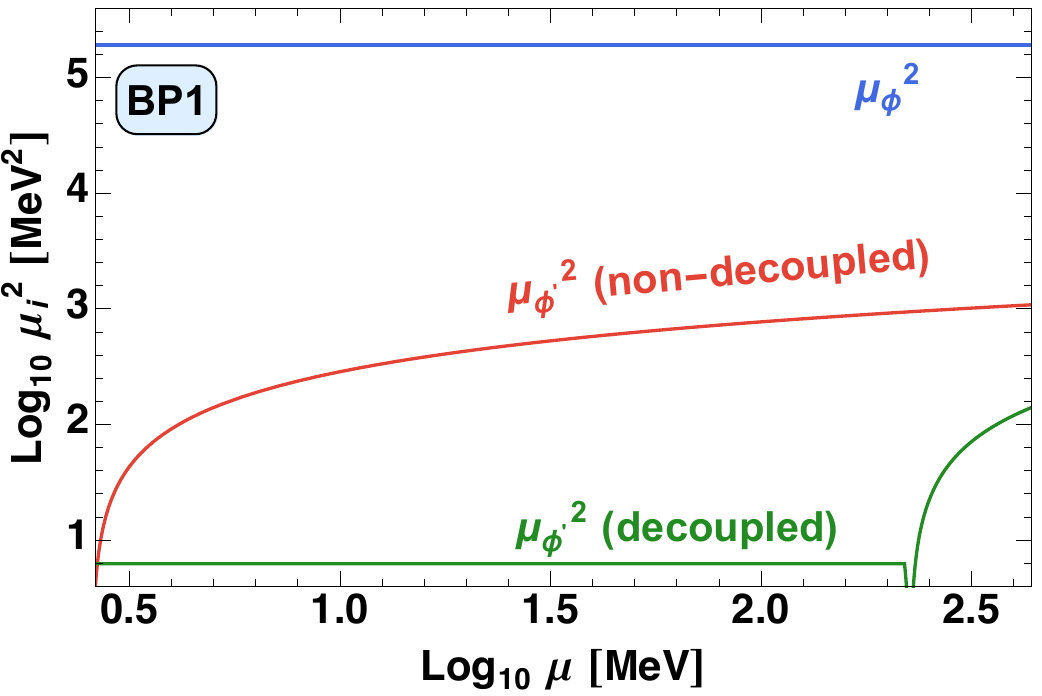}~
    \includegraphics[scale=0.4]{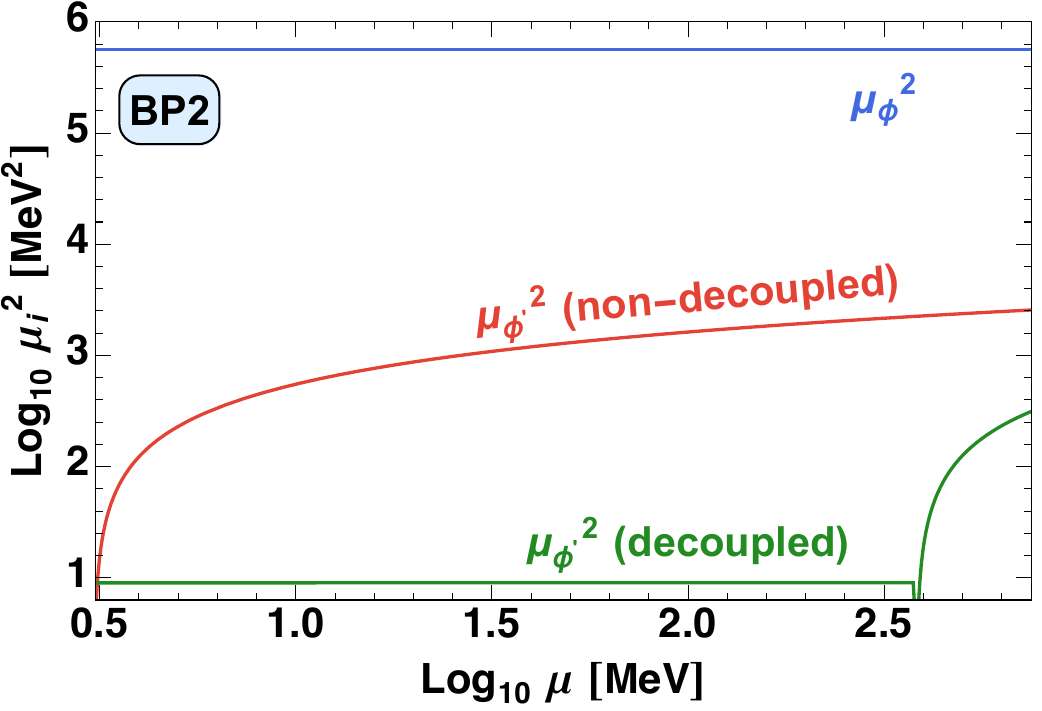}\\
    \includegraphics[scale=0.4]{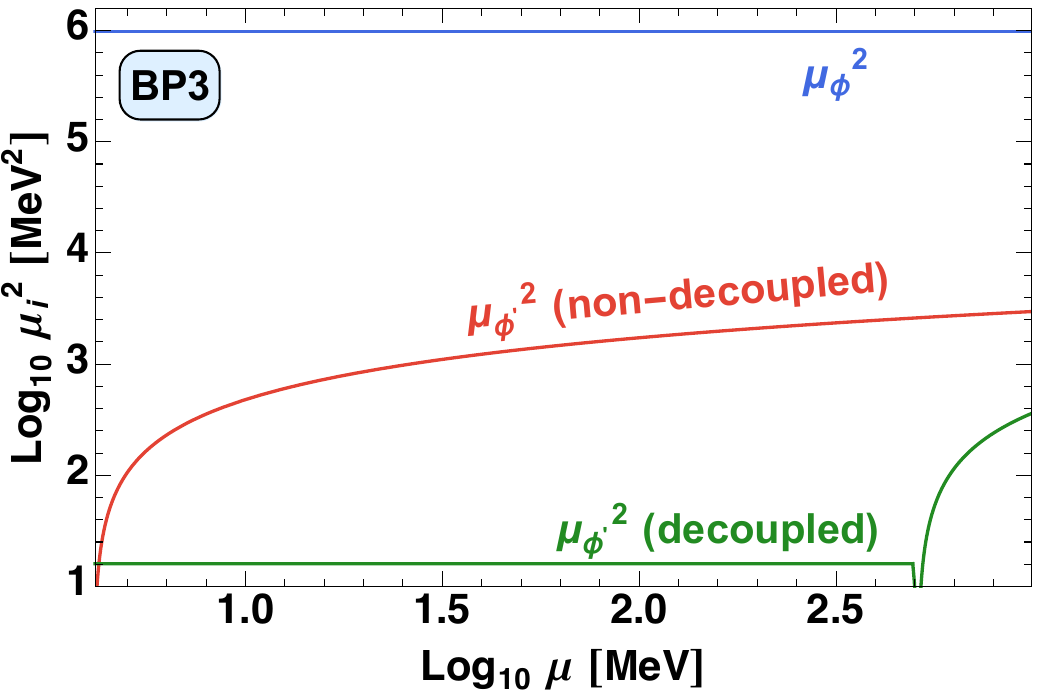}~
    \includegraphics[scale=0.4]{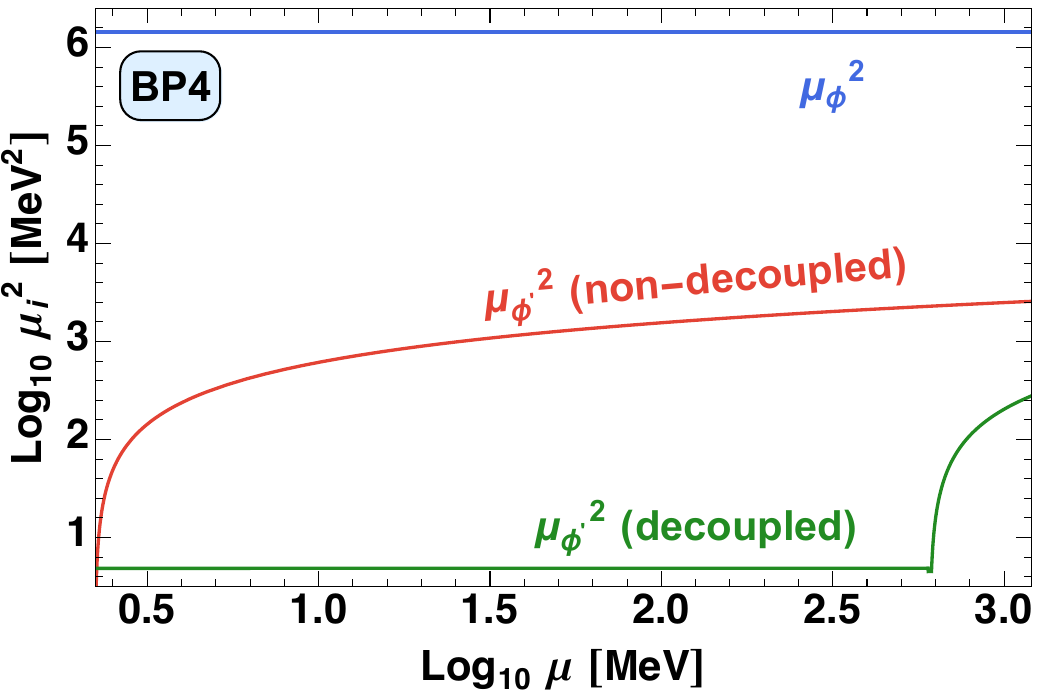}
    \caption{The running of mass parameters, $\mu_\phi^2$  with blue line, $\mu_{\phi '}^2$ with red and green line for the case of without and with decoupling, respectively.}
    \label{fig:RGE}
\end{figure*}
 
\section{Details of the decoupling method}
\label{sec:dec_method}

We utilize the decoupling method~\cite{Biondini:2020oib} to tame the effective potential from the contribution of states with mass larger than the decoupling scale $\sim m_i~e^{-3/4}$. To see the decoupling behavior of the parameters, we evaluate the following $\beta$-functions by taking the derivative of decoupled CW potential (Eq.~\eqref{VCW_deco}) with renormalization scale $\mu$,

\begin{align}
    \beta_{\mu_{\phi'}^2} = \mu \frac{\partial \mu_{\phi'}^2}{\partial \mu} &= \frac{1}{16\pi^2}(6 {\lambda'} \mu_{\phi'}^2\theta_{\phi'} + \lambda_{\phi'\phi}\mu_\phi^2\theta_\phi + 4\mu_{\phi'\phi}^2\theta_\phi) \, ,
    \label{rge:mu}
    \\
    \beta_{\lambda'} = \mu \frac{\partial \lambda'}{\partial \mu} &= \frac{1}{32\pi^2}(36{\lambda'}^2\theta_{\phi'} + \lambda_{\phi'\phi}^2\theta_{\phi}+ \lambda_{\phi'\phi_1}^2\theta_{\phi_1}) \, ,
    \label{rge:lambda}
\end{align}
where $\theta_i\equiv \theta(\tilde{\mu}-m_i)$ with $\tilde{\mu}^2=\mu^2 e^{3/2}$. The other parameters do not run as only the $\phi'$ field gets VEV in our model. In Fig.~\ref{fig:RGE}, we show the running of mass parameter with renormalization scale for our four chosen benchmark points. The running of $\mu_\phi^2$, $\mu_{\phi'}^2$ without and with decoupling, is shown with blue (constant), red and green solid line, respectively. With energy, the parameter $\mu_{\phi'}^2$ gets large correction from the heavier mass scale, $\mu_{\phi}^2$, before the decoupling is applied. After integrating out the $\phi$ field, $\mu_{\phi'}^2$ remains small till the decoupling scale. This method ensures minimal 1-loop corrections, resulting in a well-behaved effective potential at low energy scale.

\section{Gravitational waves from FOPT}
\label{appen3}
GW spectrum due to collision of bubbles
	\begin{eqnarray}
\Omega_\phi h^2 &=& 1.67 \times 10^{-5} \left ( \frac{100}{g_*} \right)^{1/3} \left(\frac{\mathcal{H}_*}{\beta}\right)^2 \left(\frac{\kappa_\phi \alpha_*}{1+\alpha_*}\right)^2 \nonumber\\&\times&\frac{A (a+b)^c}{\left[ b (f/f^\phi_{\rm peak})^{-a/c}+ a(f/f^\phi_{\rm peak})^{b/c} \right]^c} ,
	\end{eqnarray}
where, $a=1.03, b=1.84, c=1.45$ and A=5.93$\times10^{-2}$ and the peak frequency being \cite{Lewicki:2022pdb, Athron:2023xlk, Caprini:2024hue}
	\begin{equation}
 f^\phi_{\rm peak} = 1.65 \times 10^{-5} {\rm Hz} \left ( \frac{g_*}{100} \right)^{1/6} \left ( \frac{T_n}{100 \; {\rm GeV}} \right ) \frac{0.64}{2\pi} \left(\frac{\beta}{\mathcal{H}_*}\right).
	\end{equation}
 The efficiency factor $\kappa_\phi$ for bubble collision is given by \cite{Kamionkowski:1993fg}
 \begin{equation}
     \kappa_\phi =\frac{1}{1+0.715\alpha_*}\left( 0.715 \alpha_* + \frac{4}{27} \sqrt{\frac{3\alpha_*}{2}} \right).
 \end{equation}
	The GW spectrum generated from the sound wave in the plasma has been studied through large hydrodynamical simulations \cite{Hindmarsh:2017gnf} which has also been updated in several recent works \cite{Caprini:2019egz, Guo:2020grp, Athron:2023xlk}. The corresponding spectrum can be written as \cite{Athron:2023xlk} 
	\begin{eqnarray}
		\Omega_{\rm sw}h^2 &=& 2.59\times 10^{-6} \left(\frac{100}{g_*}\right)^{1/3}\left(\frac{\mathcal{H}_*}{\beta}\right) \left( \frac{\kappa_{\rm sw} \alpha_*}{1+\alpha_*}\right)^2 v_w\nonumber\\&\times&\frac{7^{3.5}(f/f^{\rm sw}_{\rm peak})^3}{(4+3(f/f^{\rm sw}_{\rm peak})^2)^{3.5}} \Upsilon.
	\end{eqnarray}
	The corresponding peak frequency is given by
	\begin{align}
		f^{\rm sw}_{\rm peak} & =8.9\times10^{-6}{\rm Hz} \left ( \frac{g_*}{100} \right)^{1/6} \frac{1}{v_w} \left ( \frac{T_n}{100 \; {\rm GeV}} \right )  \nonumber \\
  & \times \left(\frac{\beta}{\mathcal{H}_*}\right)\left (\frac{z_p}{10}\right ).
	\end{align}
 The efficiency factor for sound waves, applicable for relativistic bubble wall velocity $v_w \sim 1$ in our model, is \cite{Espinosa:2010hh}
 \begin{equation}
     \kappa_{\rm sw} =\frac{\alpha_*}{0.73+0.083 \sqrt{\alpha_*}+ \alpha_*}.
 \end{equation}
 Here, $\Upsilon=1-\frac{1}{\sqrt{1+2\tau_{\rm sw} \mathcal{H}_*}}$ is a suppression factor which depends on the lifetime of sound wave $\tau_{\rm sw}$\cite{Guo:2020grp} and it can be written as $\tau_{\rm sw}\sim R_*/\bar{U}_f$ with mean bubble separation, $R_*=(8\pi)^{1/3}v_w \beta$ and rms fluid velocity, $\bar{U}_f=\sqrt{3\kappa_{sw}\alpha_*/4(1+\alpha_*)}$; $z_p\sim 10$.
	Finally, the spectrum generated by the turbulence in the plasma is given by \cite{Caprini:2015zlo, Athron:2023xlk, Caprini:2024hue}
	\begin{eqnarray}
		\Omega_{\rm turb}h^2 &=& 3.35\times 10^{-4} \left(\frac{100}{g_*}\right)^{1/3}\left(\frac{\mathcal{H}_*}{\beta}\right) \left( \frac{\kappa_{\rm turb} \alpha_*}{1+\alpha_*}\right)^{1.5}\nonumber\\&\times   &v_w \frac{(f/f^{\rm turb}_{\rm peak})^3}{(1+f/f^{\rm turb}_{\rm peak})^{3.6}(1+8\pi f/h_*)}
	\end{eqnarray}
   with the peak frequency being \cite{Caprini:2015zlo}
	\begin{equation}
		f^{\rm turb}_{\rm peak}=2.7\times10^{-5}{\rm Hz} \left ( \frac{g_*}{100} \right)^{1/6} \frac{1}{v_w} \left ( \frac{T_n}{100 \; {\rm GeV}} \right )  \left(\frac{\beta}{{\mathcal{H}}_*}\right).
	\end{equation}
 The efficiency factor for turbulence is $\kappa_{\rm turb} \simeq 0.1 \kappa_{\rm sw}$ \cite{Caprini:2015zlo} and the  inverse Hubble time at the epoch of GW production, redshifted to today is
	\begin{equation}
		h_*=1.65\times10^{-5}{\rm Hz} \left ( \frac{g_*}{100} \right)^{1/6} \left ( \frac{T_n}{100 \; {\rm GeV}} \right ).
	\end{equation}

\bibliographystyle{JCAP}
\bibliography{refs, ref1, ref, ref4}
\end{document}